\documentstyle[12pt]{article}
\begin{document}
\renewcommand{\theequation}{\arabic{equation}}
\title{{Parameterization invariance and shape equations}
\\{of elastic axisymmetric vesicles}}
\author{R. Podgornik $^{\dagger}$\thanks{ On leave
from: Department of Theoretical Physics, J.Stefan Institute, P.O.Box
100, 61111 Ljubljana, Slovenia.}, S. Svetina $^{\ddagger}$ and
B. \v Zek\v s $^{\ddagger}$ \\
$\dagger$ Laboratory of Structural Biology, \\
Division of Computer Research and Technology, \\
National Institutes of Health, Bethesda, MD 20892 \\
$\ddagger$ Institute of Biophysics, Medical Faculty and J.Stefan
Institute,\\
University of Ljubljana, Lipi\v ceva 2, 61105 Ljubljana, Slovenia.}
\date{}
\maketitle
\begin{abstract}
The issue of different parameterizations of the axisymmetric vesicle
shape addressed by Hu Jian-Guo and Ou-Yang Zhong-Can [ Phys.Rev. E
{\bf 47} (1993) 461 ] is reassesed, especially as it transpires
through the corresponding Euler - Lagrange equations of the associated
elastic energy functional. It is argued that for regular, smooth
contours of vesicles with spherical topology, different
parameterizations of the surface are equivalent and that the
corresponding Euler - Lagrange equations are in essence the same. If,
however, one allows for discontinuous (higher) derivatives of the
contour line at the pole, the differently parameterized Euler -
Lagrange equations cease to be equivalent and describe different
physical problems. It nevertheless appears to be true that the elastic
energy corresponding to smooth contours remains a global minimum.

\end{abstract}


Recently Hu Jian-Guo and Ou-Yang Zhong-Can \cite{ref1} argued to the
effect that different parameterizations of the axisymmetric shape of a
lipid vesicle lead to different equations governing its shape. Should
this assertion be regarded as generally valid, it leads to disturbing
ambiguities in the results of the computation of equilibrium shapes of
the elastic vesicles.  Also many results derived till now
\cite{ref1a,ref2} and based on the assumption that the parameterization
of the surface does not alter its equilibrium form, would be cast in
grave doubt. It is thus of utmost importance to establish the precise
limits of the claims made in Ref.1, particularly since it is quite
clear from the general theory of the variational calculus
\cite{ref5} that these claims can not be generally valid.

A criticism of the variational methods employed in \cite{ref1} has
already been voiced by J\" ulicher and Seifert \cite{Seifert}, that
based their arguments on the incorrect way the boundary terms were
treated in the derivation of the variational equations for general
topologies.  Limiting ourselves to the spherical topology we will, in
what follows, investigate the variational shape equations for
different parameterizations and their range of validity.

Though all our formal derivations and associated discussion will be
centered on the bilayer - couple model, the modifications brought to
the formalism in the framework of the spontaneous curvature model -
the two are just different limiting cases of the generalized bilayer -
couple model - are known \cite{ref4} and do not in any respect modify
the conclusions reached below.

The formulation of the problem treated here is simple. We start with
the variational principle for an elastic vesicle under the constraint
of constant volume, $V$, constant surface area, $A$, and constant
difference between the areas of the two membrane surfaces in contact,
$\Delta A$. The equilibrium shape of the vesicle is obtained from the
first variation \cite{ref2}
\begin{equation}
\delta {\cal L} = \delta \left( {\cal F}_b - {{\textstyle{1 \over 4}}}LA -
{{\textstyle{1 \over 6}}}MV -
{{\textstyle{1 \over 2}}}N\Delta A   \right) = 0,
\label{equ-1}
\end{equation}
where the bending energy can be expanded in a series with respect to the
principal curvatures $c_1$ and $c_2$ of which the first term is of the form
\begin{equation}
{\cal F}_b = {{\textstyle{1 \over 4}}}\oint \left( c_1 + c_2 \right)^2 d\alpha
{}.
\label{equ0}
\end{equation}
In the above equation we omitted the Gaussian elastic term, as we
consider only vesicles of fixed (spherical) topology. Furthermore the elastic
energy, area, volume and area difference are all normalized with
respect to the corresponding values for the spherical shape \cite{ref2}, while
$L,M$ and $N$ are
the Lagrange multipliers corresponding to the constraints of fixed
area, volume and area difference.  Furthermore $d\alpha$ is the
normalized area element of an axisymmetric vesicle, {\sl i.e.} $\int
d\alpha = 1$.

We start by choosing the relevant variables describing the shape of an
axisymmetric vesicle. These are the angle $\psi (x)$ between the
tangent to the contour and the x - axis, and the contour line itself
$z(x)$. The $z$ - axis coincides with the rotational axis and is
perpendicular to the $x$ - axis.  The connection between the contour
line and $\psi (x)$ is
\begin{equation}
{{dz(x)}\over{dx}} = - \tan{\psi (x)},
\label{equ1}
\end{equation}
while the two principal curvatures of an axisymmetric shape have the form
\begin{equation}
c_p(x) = {{\sin{\psi (x)}}\over{x}}~~{\rm and}~~c_m(x) = \cos{\psi (x)}
{{d\psi (x)}\over{dx}}.
\label{equ1.5}
\end{equation}
The volume and the surface area difference can be obtained in the form
\begin{eqnarray}
V &=& {{\textstyle{3 \over 2}}} \oint x^2~c_p~d\alpha \nonumber \\
\Delta A &=& {{\textstyle{1 \over 2}}} \oint \left( c_p + c_m \right)~d\alpha .
\end{eqnarray}

We now analyse four different derivations of the variational shape
equations , all stemming from Eq.\ref{equ-1}, of which three are based
on axial radius, area and contour arclength parameterizations, while
the last one is a "general" shape equation
\cite{ref3}, that for an axisymmetric shape reduces to ${{\textstyle{1 \over
{c_p}}}}$
parameterization. To the analysis of \cite{ref1} that contains the
comparison of the first and the third of these parameterizations and
the "general" shape equation we thus add an additional derivation of
the shape equation, where the area itself is taken as a variational
parameter.

If we start by taking the axial radius $x$ as the parameter of the
shape, we obtain the contour line $z(x)$ from the minimization of the
action functional ${\cal L}$ which can be cast into a dimensionless
form
\begin{equation}
{\cal L} = {{\textstyle{1 \over 8}}}\int L_{x}\left( x, \psi (x),
{{d\psi (x)}\over{dx}}\right) dx,
\label{equ2}
\end{equation}
where
\begin{eqnarray}
L_{x}\left( x,\psi (x), {{d\psi (x)}\over{dx}} \right) &=& {x\over{\cos{\psi}}}
\left( {{\sin{\psi}}\over x} + \cos{\psi}{{d\psi}\over{dx}} \right)^2 -
\nonumber\\
&-& {1\over{\cos{\psi}}}\left( L x + M x^2\sin{\psi} + Nx\left( {{\sin{\psi}
}\over x} + \cos{\psi} {{d\psi}\over{dx}} \right)   \right).
\nonumber\\
{}.
\label{equ3}
\end{eqnarray}
The Euler - Lagrange equation corresponding to the x - parameterization can be
obtained from Eq.\ref{equ2} in the form:
\begin{eqnarray}
{\cal H}(x) &\equiv& \cos{^3\psi (x)} {{d^2\psi (x)}\over{dx^2}} -
{{\textstyle{1 \over 2}}}\sin{\psi (x)}
\cos{^2\psi (x)}\left( {{d\psi (x)}\over{dx}} \right)^2 + {{\cos{^3\psi (x)}}
\over x} \left( {{d\psi (x)}\over{dx}} \right) - \nonumber\\
&-&{x\over 2}\left[ 2{{\sin{\psi (x)}}
\over{x^3}} - \left( {{\sin{\psi (x)}}\over x}\right)^3 - M - L{{\sin{\psi
(x)}}
\over x} - N \left( {{\sin{\psi (x)}}\over x}\right)^2 \right] = 0.
\label{equ4}
\end{eqnarray}
One notes here that this equation is of the second order with respect
to the derivatives of $\psi (x)$. The contour line follows after a
solution of Eq.\ref{equ4} is inserted into Eq.\ref{equ1}.

We now proceed with the surface area parameterization of the vesicle
shape.  We take the axisymmetric area element in the form
\begin{equation}
d\alpha = {{\textstyle{1 \over 2}}}{{x~dx}\over{\cos{\psi (x)}}},
\label{equ5}
\end{equation}
together with $\psi = \psi ( \alpha )$.  In this parameterization the
Lagrangian is obtained as
\begin{equation}
{\cal L} = {{\textstyle{1 \over 4}}}\int L_{\alpha} \left( x, \psi (\alpha ),
{{d\psi (\alpha )}\over
{d\alpha}} \right) d\alpha
\label{equ6}
\end{equation}
with
\begin{eqnarray}
L_{\alpha} \left( x, \psi (\alpha ), {{d\psi (\alpha )}\over
{d\alpha}} \right) &=& \left( {x\over 2}{{d\psi }\over{d\alpha}} + {{\sin{\psi
}}
\over{x}}\right)^2 - L - Mx\sin{\psi } - N\left( {x\over 2}{{d\psi }\over
{d\alpha}} + {{\sin{\psi }} \over{x}}\right) + \nonumber\\
&+& \gamma_{\alpha}\left( {x\over 2}
{{dx}\over{d\alpha}} - \cos{\psi}\right).
\label{equ7}
\end{eqnarray}
The last term in the above equation is due to the constraint
Eq.\ref{equ5} in the form of a differential relation between the
surface area element and the axial radius.

In the $\alpha$ - parameterization we now obtain two Euler - Lagrange
equations, one for $\psi$ field, which is of the second order in the
derivatives of $\psi$, and the other one for $\gamma_{\alpha}$ field,
which is of the first order.  Combining these two equations in
analogous way as in Ref.1 into a single equivalent equation for the
$\psi$ field of a higher (third) order , we are led to
\begin{equation}
{{d{\cal H}(x)}\over{dx}} = 0 ~~\rightarrow~~{\cal H}(x) = const.,
\label{equ9}
\end{equation}
which is now the Euler - Lagrange equation in the $\alpha$ -
parameterization.  The ${\cal H}(x)$ function is the same as has been
defined in Eq.\ref{equ4}.

Taking the arclength of the vesicle contour as a parameter, thus
\begin{equation}
ds = {{dx}\over{\cos{\psi (x)}}},
\label{equ10}
\end{equation}
and $\psi = \psi (s)$,
we obtain the Lagrangian in the form
\begin{equation}
{\cal L} = {\textstyle{1\over 8}} \int L_{s} \left( x, \psi (s), {{d\psi
(s)}\over
{ds}} \right) ds,
\label{equ11}
\end{equation}
where
\begin{eqnarray}
L_{s} \left( x, \psi (s), {{d\psi (s)}\over
{ds}} \right) &=& x\left({{d\psi }\over{ds}} + {{\sin{\psi }}
\over{x}}\right)^2 - Lx - Mx^2\sin{\psi } - Nx\left( {{d\psi }\over
{ds}} + {{\sin{\psi }} \over{x}}\right) + \nonumber\\
&+& \gamma_s\left( {{dx}\over{ds}} - \cos{\psi}\right).
\label{equ12}
\end{eqnarray}
The last term in the above equation again ensures that the geometric
constraint Eq.\ref{equ10} is obeyed along the whole of the contour.

The Euler - Lagrange equations are once again two, one for the $\psi$
field and the other one for the $\gamma_{s}$ field. Combining the two
of them into a single higher order equation for $\psi$ we derive
\begin{equation}
{{d{\cal H}(x)}\over{dx}} + {{{\cal H}(x)}\over{x}}\left( 1 -
{{c_m(x)}\over{c_p(x)}}\right) = 0 ~~
\rightarrow~~{{{\cal H}(x)}\over{c_p(x)}} = const.,
\label{equ15}
\end{equation}
where again $c_p$ and $c_m$ are the two principal curvatures and the
function ${\cal H}(x)$ has again been defined in Eq.\ref{equ4}.

The next alternative derivation of the Euler - Lagrange equations for
the shape of the vesicle proceeds by describing the variation of the
contour line in terms of infinitesimal deformations of the shape in
the direction of the local normal to the unperturbed shape
\cite{ref3}.  For an axisymmetric shape this is equivalent to taking
${{\textstyle{1 \over {c_p}}}}$ as a parameter describing the shape. This
procedure
leads yet to a new Euler - Lagrange equation, that in the case of an
axisymmetric vesicle reduces to Eq.(7) of Ref.1. After some algebra we
were able to write the corresponding Euler - Lagrange equation in
terms of ${\cal H}(x)$, that in this case assumes the form
\begin{equation}
{{d{\cal H}(x)}\over{dx}} + {{{\cal H}(x)}\over{x}} = 0 ~~
\rightarrow~~{\cal H}(x)~x = const.
\label{equ14}
\end{equation}
The same equation has been recently derived by a somewhat different
procedure also by Zheng and Liu \cite{Zhengliu} in their analysis of
the axisymmetric shape equations.

We were thus able to show that for these four different derivations of
the Euler - Lagrange equations, Eqs.\ref{equ4},\ref{equ9},\ref{equ15}
and \ref{equ14}, giving the contour line of the axisymmetric shape,
invariably reduce to a general equation of the form $ {\cal
H}(x)~f(x) = const.$, where the form of the function $f(x)$ and the
value of the constant depend on the type of derivation.

Let us now investigate how the solutions of the different Euler -
Lagrange equations differ between themselves. This is most easily seen
if we investigate the nature of the contour line close to the pole,
{\sl i.e.} $x \to 0, s \to 0,
\alpha \to 0$ in different parameterizations. By expanding the different Euler
- Lagrange equations in the vicinity of the pole we get the following
limiting behavior. In the $x$ - parameterization the expansion of
Eq.\ref{equ4} leads to
\begin{equation}
c_p (x) \cong a + bx^2 + \dots,
\label{equ16}
\end{equation}
where the two constants are determined from the constancy of volume
($V$), surface area ($A$) and the surface area difference ($\Delta A$
in the bilayer - couple model). In the area parameterization the
corresponding limiting behavior following from Eq.\ref{equ9} is
\begin{equation}
c_p (x) \cong a + b\vert x \vert + \dots,
\label{equ17}
\end{equation}
where the constant $a$ is determined from the values of $V,A,\Delta
A$, while $b = const./3$ where the constant is the same as in
Eq.\ref{equ9}. The arclength parameterization leads to the following
limiting law
\begin{equation}
c_p (x) \cong a + b\vert x \vert + \dots,
\label{equ19}
\end{equation}
with $ {{3b}\over a} = const.$, where the constant is the same as in Eq.
\ref{equ15}.

Finally for the Euler - Lagrange equation Eq.\ref{equ14},
corresponding to locally normal shape variations introduced by Ou-Yang
and Helfrich \cite{ref3} , one gets
\begin{equation}
c_p (x) \cong a\log{\vert x \vert} + \dots,
\label{equ18}
\end{equation}
with $a = const./2$, where the constant is the same as in
Eq.\ref{equ14}. Interestingly enough an equation of the form
Eq. \ref{equ18} has been derived recently by Naito {\sl et al}.
\cite{Naito} as a general solution (not just a limiting form as is the
case in Eq. \ref{equ18}) of their shape equation, which is
just the Euler - Lagrange equation in the Ou-Yang - Helfrich
parameterization.

One can easily demonstrate that the lowest power $n$ in the expansion
$ c_p (x) \cong a + bx^n + \dots$ compatible with at least one of the
four Euler - Lagrange equations is $n = "0"$ (in the sense that
$"x^{n=0}" = \log{x}$ ), corresponding to the Ou-Yang - Helfrich
parameterization. On the other hand if we choose the constants in
Eqs.\ref{equ9}, \ref{equ15} and \ref{equ14} as zero, all the $c_p$'s
reduce to the limiting form Eq.
\ref{equ16}.

The contour line close to the poles is then obtained in an approximate
form as
\begin{equation}
{{dz(x)}\over{dx}} = -\tan{\psi (x)} \cong - \left( xc_p(x) \right).
\label{equ20}
\end{equation}
It is thus clear that only the form Eq.\ref{equ16} keeps the contour
line and all its higher derivatives finite close to the pole. The
other limiting forms lead to discontinuities in the third derivative
of the form
\begin{equation}
{{d^3z(x)}\over{dx^3}} \cong -2b~{\rm sgn}(x)
\end{equation}
for the limiting laws Eqs.\ref{equ17} and \ref{equ19}, or of the form
\begin{equation}
{{d^3z(x)}\over{dx^3}} \cong - {a\over{\vert x\vert }}~{\rm sgn}(x)
\label{discont}
\end{equation}
for the limiting case Eq.\ref{equ18}.

The existence of discontinuities in (higher) derivatives of the
contour $z(x)$ is in general connected with point source terms in the
Lagrangian. Let us investigate this connection in the case of the
Ou-Yang - Helfrich parameterization. An external force would in
general introduce a term of the form
\begin{equation}
{\cal F} = {\cal F}_b + {\textstyle{1\over 4}} \int f(x)z(x)dx
\label{external}
\end{equation}
into the energy of the vesicle, where $f(x)$ is the linear density of
the force in the direction of z - axis and the numeric coefficient has
been chosen for later convenience (if there were external momenta
acting on the contour we would in general have to include those
contributions too). As Eq.
\ref{discont} refers to a discontinuity at the pole, we presume that
the force density has the form $f(x) = f_0 \delta(x)$, where $f_0$ is
a constant. Deriving now the Euler - Lagrange equation for the {\sl
ansatz} Eq.\ref{external} we can cast it in the familiar form
\begin{equation}
{\cal H}(x) x = f_0.
\label{lad}
\end{equation}
The constant in Eq.\ref{equ14} is thus nothing but the magnitude of
the point external force acting at both poles. Instead of using the
language of point sources in the Lagrangian we could also invoke
external constraints: if the separation between the poles is kept
constant $\Delta z = const.$, then $f_0$ is just the conjugate
variable to $\Delta z$.

Either way the conclusion is that the non-analyticities of the Hu -
Ou-Yang type in the derivatives of the contour line at the poles of
the form Eq.\ref{discont} exist only if there are external forces
acting on the poles, pulling (pushing) them apart (together). More
complicated external sources (constraints) are needed in the case of
other parameterizations.

Another conclusion pertinent to the above discussion is that no matter
what the magnitude of $f_0$ might be, the corresponding membrane
bending energy will
always be larger than in the $f_0 = 0$ case. On Fig.1 we present a
result of numerical computation of ${\cal F}_b$ Eq.\ref{equ0} for a
range of $f_0$ values while keeping $V$ and $\Delta A$ constant ( $V =
0.95$ and $\Delta A = 1.0129$). Obviously $f_0 = 0$ case represents a
global minimum of this energy.

One is thus lead to the following conclusion regarding the different
forms of the Euler - Lagrange equation, stemming from the various
parameterizations of the vesicle shape. If one demands that the contour
line be a smooth function, without any discontinuities in its values
or its derivatives, then all the different Euler - Lagrange equations
reduce to a single equation of the form ${\cal H}(x) = 0$, identical
to the x - parameterization result. This is consistent with the
parameterization invariance of the variational problem, but is at odds
with the generality of the claims made in Ref.1. On the other hand if
one allows for discontinuities in the (higher) derivatives of the
contour line of the vesicle, then each Euler - Lagrange equation
corresponding to different parameterizations gives different
equilibrium shapes. However, these shapes correspond to different
physical problems. It appears that only the parameterization where
${\cal H}(x) = 0$ describes a completely unconstrained vesicle of
spherical topology. The
other parameterizations describe the shape of a vesicle under external
forces and are not relevant to the problem of the shapes of an
unconstrained vesicle, but might provide additional insight into the
shapes of constrained vesicles.

The conclusion of this paper, contrary to the assesments of
\cite{ref1}, is that for "smooth" axisymmetric shapes the way in which
one derives the shape equations matters little.  All the different
parameterizations employed in the literature are equally correct and
their corresponding variational equations are parameterization
invariant - a results of the general theory of variational calculus of
long standing \cite{ref5}.  Should one, however, want to include
'pathological' shapes with various discontinuities in the higher
derivatives of the contour line one should first of all start with a
more elaborate form of the curvature energy including the external
forces or geometrical constraints that lead in each particular case to
these discontinuities. For these cases no general form of the free
energy is, however, presently in sight and the discussion of the
variational equations for those cases should reflect the explicit
constraints for each particular case.



\vskip 2 cm
\noindent
{\Large \bf Figure Captions}

\vskip 1 cm
\noindent
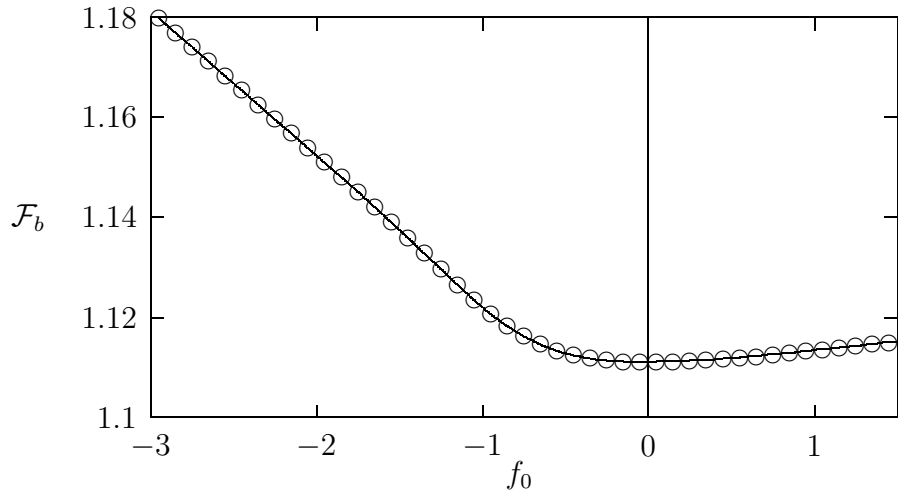
{\bf Fig.1:} Membrane bending energy ${\cal F}_b$ obtained from numerical
integration of Eq.\ref{lad} for different values of the force
$f_0$. The normalized volume is $V = 0.95$ and the normalized area
difference is $\Delta A = 1.0129$.

\begin{figure}[h]
\setlength{\unitlength}{0.240900pt}
\ifx\plotpoint\undefined\newsavebox{\plotpoint}\fi
\sbox{\plotpoint}{\rule[-0.175pt]{0.350pt}{0.350pt}}%
\begin{picture}(1500,900)(0,0)
\tenrm
\sbox{\plotpoint}{\rule[-0.175pt]{0.350pt}{0.350pt}}%
\put(1045,158){\rule[-0.175pt]{0.350pt}{151.526pt}}
\put(264,158){\rule[-0.175pt]{4.818pt}{0.350pt}}
\put(242,158){\makebox(0,0)[r]{$1.1$}}
\put(1416,158){\rule[-0.175pt]{4.818pt}{0.350pt}}
\put(264,315){\rule[-0.175pt]{4.818pt}{0.350pt}}
\put(242,315){\makebox(0,0)[r]{$1.12$}}
\put(1416,315){\rule[-0.175pt]{4.818pt}{0.350pt}}
\put(264,473){\rule[-0.175pt]{4.818pt}{0.350pt}}
\put(242,473){\makebox(0,0)[r]{$1.14$}}
\put(1416,473){\rule[-0.175pt]{4.818pt}{0.350pt}}
\put(264,630){\rule[-0.175pt]{4.818pt}{0.350pt}}
\put(242,630){\makebox(0,0)[r]{$1.16$}}
\put(1416,630){\rule[-0.175pt]{4.818pt}{0.350pt}}
\put(264,787){\rule[-0.175pt]{4.818pt}{0.350pt}}
\put(242,787){\makebox(0,0)[r]{$1.18$}}
\put(1416,787){\rule[-0.175pt]{4.818pt}{0.350pt}}
\put(264,158){\rule[-0.175pt]{0.350pt}{4.818pt}}
\put(264,113){\makebox(0,0){$-3$}}
\put(264,767){\rule[-0.175pt]{0.350pt}{4.818pt}}
\put(524,158){\rule[-0.175pt]{0.350pt}{4.818pt}}
\put(524,113){\makebox(0,0){$-2$}}
\put(524,767){\rule[-0.175pt]{0.350pt}{4.818pt}}
\put(785,158){\rule[-0.175pt]{0.350pt}{4.818pt}}
\put(785,113){\makebox(0,0){$-1$}}
\put(785,767){\rule[-0.175pt]{0.350pt}{4.818pt}}
\put(1045,158){\rule[-0.175pt]{0.350pt}{4.818pt}}
\put(1045,113){\makebox(0,0){$0$}}
\put(1045,767){\rule[-0.175pt]{0.350pt}{4.818pt}}
\put(1306,158){\rule[-0.175pt]{0.350pt}{4.818pt}}
\put(1306,113){\makebox(0,0){$1$}}
\put(1306,767){\rule[-0.175pt]{0.350pt}{4.818pt}}
\put(264,158){\rule[-0.175pt]{282.335pt}{0.350pt}}
\put(1436,158){\rule[-0.175pt]{0.350pt}{151.526pt}}
\put(264,787){\rule[-0.175pt]{282.335pt}{0.350pt}}
\put(45,472){\makebox(0,0)[l]{\shortstack{${\cal F}_b$ }}}
\put(850,68){\makebox(0,0){$f_0$ }}
\put(264,158){\rule[-0.175pt]{0.350pt}{151.526pt}}
\put(1423,277){\rule[-0.175pt]{3.132pt}{0.350pt}}
\put(1414,276){\rule[-0.175pt]{2.088pt}{0.350pt}}
\put(1405,275){\rule[-0.175pt]{2.088pt}{0.350pt}}
\put(1397,274){\rule[-0.175pt]{2.088pt}{0.350pt}}
\put(1388,273){\rule[-0.175pt]{2.088pt}{0.350pt}}
\put(1379,272){\rule[-0.175pt]{2.088pt}{0.350pt}}
\put(1371,271){\rule[-0.175pt]{2.088pt}{0.350pt}}
\put(1362,270){\rule[-0.175pt]{2.088pt}{0.350pt}}
\put(1353,269){\rule[-0.175pt]{2.088pt}{0.350pt}}
\put(1345,268){\rule[-0.175pt]{2.088pt}{0.350pt}}
\put(1332,267){\rule[-0.175pt]{3.132pt}{0.350pt}}
\put(1319,266){\rule[-0.175pt]{3.132pt}{0.350pt}}
\put(1310,265){\rule[-0.175pt]{2.088pt}{0.350pt}}
\put(1301,264){\rule[-0.175pt]{2.088pt}{0.350pt}}
\put(1293,263){\rule[-0.175pt]{2.088pt}{0.350pt}}
\put(1284,262){\rule[-0.175pt]{2.088pt}{0.350pt}}
\put(1275,261){\rule[-0.175pt]{2.088pt}{0.350pt}}
\put(1267,260){\rule[-0.175pt]{2.088pt}{0.350pt}}
\put(1254,259){\rule[-0.175pt]{3.132pt}{0.350pt}}
\put(1241,258){\rule[-0.175pt]{3.132pt}{0.350pt}}
\put(1232,257){\rule[-0.175pt]{2.088pt}{0.350pt}}
\put(1223,256){\rule[-0.175pt]{2.088pt}{0.350pt}}
\put(1215,255){\rule[-0.175pt]{2.088pt}{0.350pt}}
\put(1202,254){\rule[-0.175pt]{3.132pt}{0.350pt}}
\put(1189,253){\rule[-0.175pt]{3.132pt}{0.350pt}}
\put(1176,252){\rule[-0.175pt]{3.132pt}{0.350pt}}
\put(1163,251){\rule[-0.175pt]{3.132pt}{0.350pt}}
\put(1149,250){\rule[-0.175pt]{3.252pt}{0.350pt}}
\put(1136,249){\rule[-0.175pt]{3.252pt}{0.350pt}}
\put(1110,248){\rule[-0.175pt]{6.263pt}{0.350pt}}
\put(1084,247){\rule[-0.175pt]{6.263pt}{0.350pt}}
\put(1058,246){\rule[-0.175pt]{6.263pt}{0.350pt}}
\put(1006,245){\rule[-0.175pt]{12.527pt}{0.350pt}}
\put(993,246){\rule[-0.175pt]{3.132pt}{0.350pt}}
\put(980,247){\rule[-0.175pt]{3.132pt}{0.350pt}}
\put(971,248){\rule[-0.175pt]{2.088pt}{0.350pt}}
\put(962,249){\rule[-0.175pt]{2.088pt}{0.350pt}}
\put(954,250){\rule[-0.175pt]{2.088pt}{0.350pt}}
\put(948,251){\rule[-0.175pt]{1.253pt}{0.350pt}}
\put(943,252){\rule[-0.175pt]{1.253pt}{0.350pt}}
\put(938,253){\rule[-0.175pt]{1.253pt}{0.350pt}}
\put(933,254){\rule[-0.175pt]{1.253pt}{0.350pt}}
\put(928,255){\rule[-0.175pt]{1.253pt}{0.350pt}}
\put(924,256){\rule[-0.175pt]{0.895pt}{0.350pt}}
\put(920,257){\rule[-0.175pt]{0.895pt}{0.350pt}}
\put(916,258){\rule[-0.175pt]{0.895pt}{0.350pt}}
\put(913,259){\rule[-0.175pt]{0.895pt}{0.350pt}}
\put(909,260){\rule[-0.175pt]{0.895pt}{0.350pt}}
\put(905,261){\rule[-0.175pt]{0.895pt}{0.350pt}}
\put(902,262){\rule[-0.175pt]{0.895pt}{0.350pt}}
\put(899,263){\rule[-0.175pt]{0.626pt}{0.350pt}}
\put(896,264){\rule[-0.175pt]{0.626pt}{0.350pt}}
\put(894,265){\rule[-0.175pt]{0.626pt}{0.350pt}}
\put(891,266){\rule[-0.175pt]{0.626pt}{0.350pt}}
\put(889,267){\rule[-0.175pt]{0.626pt}{0.350pt}}
\put(886,268){\rule[-0.175pt]{0.626pt}{0.350pt}}
\put(883,269){\rule[-0.175pt]{0.626pt}{0.350pt}}
\put(881,270){\rule[-0.175pt]{0.626pt}{0.350pt}}
\put(878,271){\rule[-0.175pt]{0.626pt}{0.350pt}}
\put(876,272){\rule[-0.175pt]{0.626pt}{0.350pt}}
\put(874,273){\rule[-0.175pt]{0.482pt}{0.350pt}}
\put(872,274){\rule[-0.175pt]{0.482pt}{0.350pt}}
\put(870,275){\rule[-0.175pt]{0.482pt}{0.350pt}}
\put(868,276){\rule[-0.175pt]{0.482pt}{0.350pt}}
\put(866,277){\rule[-0.175pt]{0.482pt}{0.350pt}}
\put(864,278){\rule[-0.175pt]{0.482pt}{0.350pt}}
\put(862,279){\rule[-0.175pt]{0.482pt}{0.350pt}}
\put(860,280){\rule[-0.175pt]{0.482pt}{0.350pt}}
\put(858,281){\rule[-0.175pt]{0.482pt}{0.350pt}}
\put(856,282){\rule[-0.175pt]{0.482pt}{0.350pt}}
\put(854,283){\rule[-0.175pt]{0.482pt}{0.350pt}}
\put(852,284){\rule[-0.175pt]{0.482pt}{0.350pt}}
\put(850,285){\rule[-0.175pt]{0.482pt}{0.350pt}}
\put(848,286){\rule[-0.175pt]{0.391pt}{0.350pt}}
\put(846,287){\rule[-0.175pt]{0.391pt}{0.350pt}}
\put(845,288){\rule[-0.175pt]{0.391pt}{0.350pt}}
\put(843,289){\rule[-0.175pt]{0.391pt}{0.350pt}}
\put(841,290){\rule[-0.175pt]{0.391pt}{0.350pt}}
\put(840,291){\rule[-0.175pt]{0.391pt}{0.350pt}}
\put(838,292){\rule[-0.175pt]{0.391pt}{0.350pt}}
\put(837,293){\rule[-0.175pt]{0.391pt}{0.350pt}}
\put(835,294){\rule[-0.175pt]{0.391pt}{0.350pt}}
\put(833,295){\rule[-0.175pt]{0.391pt}{0.350pt}}
\put(832,296){\rule[-0.175pt]{0.391pt}{0.350pt}}
\put(830,297){\rule[-0.175pt]{0.391pt}{0.350pt}}
\put(828,298){\rule[-0.175pt]{0.391pt}{0.350pt}}
\put(827,299){\rule[-0.175pt]{0.391pt}{0.350pt}}
\put(825,300){\rule[-0.175pt]{0.391pt}{0.350pt}}
\put(824,301){\rule[-0.175pt]{0.391pt}{0.350pt}}
\put(822,302){\usebox{\plotpoint}}
\put(821,303){\usebox{\plotpoint}}
\put(819,304){\usebox{\plotpoint}}
\put(818,305){\usebox{\plotpoint}}
\put(817,306){\usebox{\plotpoint}}
\put(815,307){\usebox{\plotpoint}}
\put(814,308){\usebox{\plotpoint}}
\put(813,309){\usebox{\plotpoint}}
\put(811,310){\usebox{\plotpoint}}
\put(810,311){\usebox{\plotpoint}}
\put(808,312){\usebox{\plotpoint}}
\put(807,313){\usebox{\plotpoint}}
\put(806,314){\usebox{\plotpoint}}
\put(804,315){\usebox{\plotpoint}}
\put(803,316){\usebox{\plotpoint}}
\put(802,317){\usebox{\plotpoint}}
\put(800,318){\usebox{\plotpoint}}
\put(799,319){\usebox{\plotpoint}}
\put(798,320){\usebox{\plotpoint}}
\put(796,321){\usebox{\plotpoint}}
\put(795,322){\usebox{\plotpoint}}
\put(794,323){\usebox{\plotpoint}}
\put(793,324){\usebox{\plotpoint}}
\put(792,325){\usebox{\plotpoint}}
\put(790,326){\usebox{\plotpoint}}
\put(789,327){\usebox{\plotpoint}}
\put(788,328){\usebox{\plotpoint}}
\put(787,329){\usebox{\plotpoint}}
\put(786,330){\usebox{\plotpoint}}
\put(784,331){\usebox{\plotpoint}}
\put(783,332){\usebox{\plotpoint}}
\put(782,333){\usebox{\plotpoint}}
\put(781,334){\usebox{\plotpoint}}
\put(780,335){\usebox{\plotpoint}}
\put(779,336){\usebox{\plotpoint}}
\put(777,337){\usebox{\plotpoint}}
\put(776,338){\usebox{\plotpoint}}
\put(775,339){\usebox{\plotpoint}}
\put(774,340){\usebox{\plotpoint}}
\put(773,341){\usebox{\plotpoint}}
\put(772,342){\usebox{\plotpoint}}
\put(770,343){\usebox{\plotpoint}}
\put(769,344){\usebox{\plotpoint}}
\put(768,345){\usebox{\plotpoint}}
\put(767,346){\usebox{\plotpoint}}
\put(766,347){\usebox{\plotpoint}}
\put(765,348){\usebox{\plotpoint}}
\put(764,349){\usebox{\plotpoint}}
\put(763,350){\usebox{\plotpoint}}
\put(762,351){\usebox{\plotpoint}}
\put(761,352){\usebox{\plotpoint}}
\put(760,353){\usebox{\plotpoint}}
\put(759,354){\usebox{\plotpoint}}
\put(757,355){\usebox{\plotpoint}}
\put(756,356){\usebox{\plotpoint}}
\put(755,357){\usebox{\plotpoint}}
\put(754,358){\usebox{\plotpoint}}
\put(753,359){\usebox{\plotpoint}}
\put(752,360){\usebox{\plotpoint}}
\put(751,361){\usebox{\plotpoint}}
\put(750,362){\usebox{\plotpoint}}
\put(749,363){\usebox{\plotpoint}}
\put(748,364){\usebox{\plotpoint}}
\put(747,365){\usebox{\plotpoint}}
\put(746,366){\usebox{\plotpoint}}
\put(744,367){\usebox{\plotpoint}}
\put(743,368){\usebox{\plotpoint}}
\put(742,369){\usebox{\plotpoint}}
\put(741,370){\usebox{\plotpoint}}
\put(740,371){\usebox{\plotpoint}}
\put(739,372){\usebox{\plotpoint}}
\put(738,373){\usebox{\plotpoint}}
\put(737,374){\usebox{\plotpoint}}
\put(736,375){\usebox{\plotpoint}}
\put(735,376){\usebox{\plotpoint}}
\put(734,377){\usebox{\plotpoint}}
\put(733,378){\usebox{\plotpoint}}
\put(732,379){\usebox{\plotpoint}}
\put(731,380){\usebox{\plotpoint}}
\put(730,381){\usebox{\plotpoint}}
\put(729,382){\usebox{\plotpoint}}
\put(728,383){\usebox{\plotpoint}}
\put(727,384){\usebox{\plotpoint}}
\put(726,385){\usebox{\plotpoint}}
\put(725,386){\usebox{\plotpoint}}
\put(724,387){\usebox{\plotpoint}}
\put(723,388){\usebox{\plotpoint}}
\put(722,389){\usebox{\plotpoint}}
\put(721,390){\usebox{\plotpoint}}
\put(720,391){\usebox{\plotpoint}}
\put(718,392){\usebox{\plotpoint}}
\put(717,393){\usebox{\plotpoint}}
\put(716,394){\usebox{\plotpoint}}
\put(715,395){\usebox{\plotpoint}}
\put(714,396){\usebox{\plotpoint}}
\put(713,397){\usebox{\plotpoint}}
\put(712,398){\usebox{\plotpoint}}
\put(711,399){\usebox{\plotpoint}}
\put(710,400){\usebox{\plotpoint}}
\put(709,401){\usebox{\plotpoint}}
\put(708,402){\usebox{\plotpoint}}
\put(707,403){\usebox{\plotpoint}}
\put(705,404){\usebox{\plotpoint}}
\put(704,405){\usebox{\plotpoint}}
\put(703,406){\usebox{\plotpoint}}
\put(702,407){\usebox{\plotpoint}}
\put(701,408){\usebox{\plotpoint}}
\put(700,409){\usebox{\plotpoint}}
\put(699,410){\usebox{\plotpoint}}
\put(698,411){\usebox{\plotpoint}}
\put(697,412){\usebox{\plotpoint}}
\put(696,413){\usebox{\plotpoint}}
\put(695,414){\usebox{\plotpoint}}
\put(694,415){\usebox{\plotpoint}}
\put(692,416){\usebox{\plotpoint}}
\put(691,417){\usebox{\plotpoint}}
\put(690,418){\usebox{\plotpoint}}
\put(689,419){\usebox{\plotpoint}}
\put(688,420){\usebox{\plotpoint}}
\put(687,421){\usebox{\plotpoint}}
\put(686,422){\usebox{\plotpoint}}
\put(685,423){\usebox{\plotpoint}}
\put(684,424){\usebox{\plotpoint}}
\put(683,425){\usebox{\plotpoint}}
\put(682,426){\usebox{\plotpoint}}
\put(681,427){\usebox{\plotpoint}}
\put(680,428){\usebox{\plotpoint}}
\put(679,429){\usebox{\plotpoint}}
\put(678,430){\usebox{\plotpoint}}
\put(677,431){\usebox{\plotpoint}}
\put(676,432){\usebox{\plotpoint}}
\put(675,433){\usebox{\plotpoint}}
\put(674,434){\usebox{\plotpoint}}
\put(673,435){\usebox{\plotpoint}}
\put(672,436){\usebox{\plotpoint}}
\put(671,437){\usebox{\plotpoint}}
\put(670,438){\usebox{\plotpoint}}
\put(669,439){\usebox{\plotpoint}}
\put(668,440){\usebox{\plotpoint}}
\put(666,441){\usebox{\plotpoint}}
\put(665,442){\usebox{\plotpoint}}
\put(664,443){\usebox{\plotpoint}}
\put(663,444){\usebox{\plotpoint}}
\put(662,445){\usebox{\plotpoint}}
\put(661,446){\usebox{\plotpoint}}
\put(660,447){\usebox{\plotpoint}}
\put(659,448){\usebox{\plotpoint}}
\put(658,449){\usebox{\plotpoint}}
\put(657,450){\usebox{\plotpoint}}
\put(656,451){\usebox{\plotpoint}}
\put(655,452){\usebox{\plotpoint}}
\put(653,453){\usebox{\plotpoint}}
\put(652,454){\usebox{\plotpoint}}
\put(651,455){\usebox{\plotpoint}}
\put(650,456){\usebox{\plotpoint}}
\put(649,457){\usebox{\plotpoint}}
\put(648,458){\usebox{\plotpoint}}
\put(647,459){\usebox{\plotpoint}}
\put(646,460){\usebox{\plotpoint}}
\put(645,461){\usebox{\plotpoint}}
\put(644,462){\usebox{\plotpoint}}
\put(643,463){\usebox{\plotpoint}}
\put(642,464){\usebox{\plotpoint}}
\put(640,465){\usebox{\plotpoint}}
\put(639,466){\usebox{\plotpoint}}
\put(638,467){\usebox{\plotpoint}}
\put(637,468){\usebox{\plotpoint}}
\put(636,469){\usebox{\plotpoint}}
\put(635,470){\usebox{\plotpoint}}
\put(634,471){\usebox{\plotpoint}}
\put(633,472){\usebox{\plotpoint}}
\put(632,473){\usebox{\plotpoint}}
\put(631,474){\usebox{\plotpoint}}
\put(630,475){\usebox{\plotpoint}}
\put(629,476){\usebox{\plotpoint}}
\put(627,477){\usebox{\plotpoint}}
\put(626,478){\usebox{\plotpoint}}
\put(625,479){\usebox{\plotpoint}}
\put(624,480){\usebox{\plotpoint}}
\put(623,481){\usebox{\plotpoint}}
\put(622,482){\usebox{\plotpoint}}
\put(621,483){\usebox{\plotpoint}}
\put(620,484){\usebox{\plotpoint}}
\put(619,485){\usebox{\plotpoint}}
\put(618,486){\usebox{\plotpoint}}
\put(617,487){\usebox{\plotpoint}}
\put(616,488){\usebox{\plotpoint}}
\put(614,489){\usebox{\plotpoint}}
\put(613,490){\usebox{\plotpoint}}
\put(612,491){\usebox{\plotpoint}}
\put(611,492){\usebox{\plotpoint}}
\put(610,493){\usebox{\plotpoint}}
\put(609,494){\usebox{\plotpoint}}
\put(608,495){\usebox{\plotpoint}}
\put(607,496){\usebox{\plotpoint}}
\put(606,497){\usebox{\plotpoint}}
\put(605,498){\usebox{\plotpoint}}
\put(604,499){\usebox{\plotpoint}}
\put(603,500){\usebox{\plotpoint}}
\put(601,501){\usebox{\plotpoint}}
\put(600,502){\usebox{\plotpoint}}
\put(599,503){\usebox{\plotpoint}}
\put(598,504){\usebox{\plotpoint}}
\put(597,505){\usebox{\plotpoint}}
\put(596,506){\usebox{\plotpoint}}
\put(595,507){\usebox{\plotpoint}}
\put(594,508){\usebox{\plotpoint}}
\put(593,509){\usebox{\plotpoint}}
\put(592,510){\usebox{\plotpoint}}
\put(591,511){\usebox{\plotpoint}}
\put(590,512){\usebox{\plotpoint}}
\put(588,513){\usebox{\plotpoint}}
\put(587,514){\usebox{\plotpoint}}
\put(586,515){\usebox{\plotpoint}}
\put(585,516){\usebox{\plotpoint}}
\put(584,517){\usebox{\plotpoint}}
\put(583,518){\usebox{\plotpoint}}
\put(582,519){\usebox{\plotpoint}}
\put(580,520){\usebox{\plotpoint}}
\put(579,521){\usebox{\plotpoint}}
\put(578,522){\usebox{\plotpoint}}
\put(577,523){\usebox{\plotpoint}}
\put(576,524){\usebox{\plotpoint}}
\put(575,525){\usebox{\plotpoint}}
\put(574,526){\usebox{\plotpoint}}
\put(573,527){\usebox{\plotpoint}}
\put(571,528){\usebox{\plotpoint}}
\put(570,529){\usebox{\plotpoint}}
\put(569,530){\usebox{\plotpoint}}
\put(568,531){\usebox{\plotpoint}}
\put(567,532){\usebox{\plotpoint}}
\put(566,533){\usebox{\plotpoint}}
\put(565,534){\usebox{\plotpoint}}
\put(564,535){\usebox{\plotpoint}}
\put(562,536){\usebox{\plotpoint}}
\put(561,537){\usebox{\plotpoint}}
\put(560,538){\usebox{\plotpoint}}
\put(559,539){\usebox{\plotpoint}}
\put(558,540){\usebox{\plotpoint}}
\put(556,541){\usebox{\plotpoint}}
\put(555,542){\usebox{\plotpoint}}
\put(554,543){\usebox{\plotpoint}}
\put(553,544){\usebox{\plotpoint}}
\put(552,545){\usebox{\plotpoint}}
\put(551,546){\usebox{\plotpoint}}
\put(549,547){\usebox{\plotpoint}}
\put(548,548){\usebox{\plotpoint}}
\put(547,549){\usebox{\plotpoint}}
\put(546,550){\usebox{\plotpoint}}
\put(545,551){\usebox{\plotpoint}}
\put(544,552){\usebox{\plotpoint}}
\put(542,553){\usebox{\plotpoint}}
\put(541,554){\usebox{\plotpoint}}
\put(540,555){\usebox{\plotpoint}}
\put(539,556){\usebox{\plotpoint}}
\put(538,557){\usebox{\plotpoint}}
\put(537,558){\usebox{\plotpoint}}
\put(535,559){\usebox{\plotpoint}}
\put(534,560){\usebox{\plotpoint}}
\put(533,561){\usebox{\plotpoint}}
\put(532,562){\usebox{\plotpoint}}
\put(531,563){\usebox{\plotpoint}}
\put(530,564){\usebox{\plotpoint}}
\put(529,565){\usebox{\plotpoint}}
\put(527,566){\usebox{\plotpoint}}
\put(526,567){\usebox{\plotpoint}}
\put(525,568){\usebox{\plotpoint}}
\put(524,569){\usebox{\plotpoint}}
\put(523,570){\usebox{\plotpoint}}
\put(522,571){\usebox{\plotpoint}}
\put(521,572){\usebox{\plotpoint}}
\put(520,573){\usebox{\plotpoint}}
\put(518,574){\usebox{\plotpoint}}
\put(517,575){\usebox{\plotpoint}}
\put(516,576){\usebox{\plotpoint}}
\put(515,577){\usebox{\plotpoint}}
\put(514,578){\usebox{\plotpoint}}
\put(513,579){\usebox{\plotpoint}}
\put(512,580){\usebox{\plotpoint}}
\put(511,581){\usebox{\plotpoint}}
\put(509,582){\usebox{\plotpoint}}
\put(508,583){\usebox{\plotpoint}}
\put(507,584){\usebox{\plotpoint}}
\put(506,585){\usebox{\plotpoint}}
\put(505,586){\usebox{\plotpoint}}
\put(504,587){\usebox{\plotpoint}}
\put(503,588){\usebox{\plotpoint}}
\put(501,589){\usebox{\plotpoint}}
\put(500,590){\usebox{\plotpoint}}
\put(499,591){\usebox{\plotpoint}}
\put(498,592){\usebox{\plotpoint}}
\put(497,593){\usebox{\plotpoint}}
\put(496,594){\usebox{\plotpoint}}
\put(495,595){\usebox{\plotpoint}}
\put(494,596){\usebox{\plotpoint}}
\put(492,597){\usebox{\plotpoint}}
\put(491,598){\usebox{\plotpoint}}
\put(490,599){\usebox{\plotpoint}}
\put(489,600){\usebox{\plotpoint}}
\put(488,601){\usebox{\plotpoint}}
\put(487,602){\usebox{\plotpoint}}
\put(486,603){\usebox{\plotpoint}}
\put(485,604){\usebox{\plotpoint}}
\put(483,605){\usebox{\plotpoint}}
\put(482,606){\usebox{\plotpoint}}
\put(481,607){\usebox{\plotpoint}}
\put(480,608){\usebox{\plotpoint}}
\put(479,609){\usebox{\plotpoint}}
\put(477,610){\usebox{\plotpoint}}
\put(476,611){\usebox{\plotpoint}}
\put(475,612){\usebox{\plotpoint}}
\put(474,613){\usebox{\plotpoint}}
\put(473,614){\usebox{\plotpoint}}
\put(471,615){\usebox{\plotpoint}}
\put(470,616){\usebox{\plotpoint}}
\put(469,617){\usebox{\plotpoint}}
\put(468,618){\usebox{\plotpoint}}
\put(467,619){\usebox{\plotpoint}}
\put(466,620){\usebox{\plotpoint}}
\put(464,621){\usebox{\plotpoint}}
\put(463,622){\usebox{\plotpoint}}
\put(462,623){\usebox{\plotpoint}}
\put(461,624){\usebox{\plotpoint}}
\put(460,625){\usebox{\plotpoint}}
\put(459,626){\usebox{\plotpoint}}
\put(457,627){\usebox{\plotpoint}}
\put(456,628){\usebox{\plotpoint}}
\put(455,629){\usebox{\plotpoint}}
\put(454,630){\usebox{\plotpoint}}
\put(453,631){\usebox{\plotpoint}}
\put(452,632){\usebox{\plotpoint}}
\put(451,633){\usebox{\plotpoint}}
\put(449,634){\usebox{\plotpoint}}
\put(448,635){\usebox{\plotpoint}}
\put(447,636){\usebox{\plotpoint}}
\put(446,637){\usebox{\plotpoint}}
\put(445,638){\usebox{\plotpoint}}
\put(444,639){\usebox{\plotpoint}}
\put(443,640){\usebox{\plotpoint}}
\put(442,641){\usebox{\plotpoint}}
\put(440,642){\usebox{\plotpoint}}
\put(439,643){\usebox{\plotpoint}}
\put(438,644){\usebox{\plotpoint}}
\put(437,645){\usebox{\plotpoint}}
\put(436,646){\usebox{\plotpoint}}
\put(435,647){\usebox{\plotpoint}}
\put(434,648){\usebox{\plotpoint}}
\put(433,649){\usebox{\plotpoint}}
\put(431,650){\usebox{\plotpoint}}
\put(430,651){\usebox{\plotpoint}}
\put(429,652){\usebox{\plotpoint}}
\put(428,653){\usebox{\plotpoint}}
\put(427,654){\usebox{\plotpoint}}
\put(426,655){\usebox{\plotpoint}}
\put(425,656){\usebox{\plotpoint}}
\put(423,657){\usebox{\plotpoint}}
\put(422,658){\usebox{\plotpoint}}
\put(421,659){\usebox{\plotpoint}}
\put(420,660){\usebox{\plotpoint}}
\put(419,661){\usebox{\plotpoint}}
\put(418,662){\usebox{\plotpoint}}
\put(417,663){\usebox{\plotpoint}}
\put(416,664){\usebox{\plotpoint}}
\put(414,665){\usebox{\plotpoint}}
\put(413,666){\usebox{\plotpoint}}
\put(412,667){\usebox{\plotpoint}}
\put(411,668){\usebox{\plotpoint}}
\put(410,669){\usebox{\plotpoint}}
\put(409,670){\usebox{\plotpoint}}
\put(408,671){\usebox{\plotpoint}}
\put(407,672){\usebox{\plotpoint}}
\put(405,673){\usebox{\plotpoint}}
\put(404,674){\usebox{\plotpoint}}
\put(403,675){\usebox{\plotpoint}}
\put(402,676){\usebox{\plotpoint}}
\put(401,677){\usebox{\plotpoint}}
\put(399,678){\usebox{\plotpoint}}
\put(398,679){\usebox{\plotpoint}}
\put(397,680){\usebox{\plotpoint}}
\put(396,681){\usebox{\plotpoint}}
\put(395,682){\usebox{\plotpoint}}
\put(393,683){\usebox{\plotpoint}}
\put(392,684){\usebox{\plotpoint}}
\put(391,685){\usebox{\plotpoint}}
\put(390,686){\usebox{\plotpoint}}
\put(389,687){\usebox{\plotpoint}}
\put(388,688){\usebox{\plotpoint}}
\put(386,689){\usebox{\plotpoint}}
\put(385,690){\usebox{\plotpoint}}
\put(384,691){\usebox{\plotpoint}}
\put(383,692){\usebox{\plotpoint}}
\put(382,693){\usebox{\plotpoint}}
\put(381,694){\usebox{\plotpoint}}
\put(379,695){\usebox{\plotpoint}}
\put(378,696){\usebox{\plotpoint}}
\put(377,697){\usebox{\plotpoint}}
\put(376,698){\usebox{\plotpoint}}
\put(375,699){\usebox{\plotpoint}}
\put(374,700){\usebox{\plotpoint}}
\put(373,701){\usebox{\plotpoint}}
\put(371,702){\usebox{\plotpoint}}
\put(370,703){\usebox{\plotpoint}}
\put(369,704){\usebox{\plotpoint}}
\put(368,705){\usebox{\plotpoint}}
\put(367,706){\usebox{\plotpoint}}
\put(366,707){\usebox{\plotpoint}}
\put(365,708){\usebox{\plotpoint}}
\put(364,709){\usebox{\plotpoint}}
\put(362,710){\usebox{\plotpoint}}
\put(361,711){\usebox{\plotpoint}}
\put(360,712){\usebox{\plotpoint}}
\put(359,713){\usebox{\plotpoint}}
\put(358,714){\usebox{\plotpoint}}
\put(357,715){\usebox{\plotpoint}}
\put(356,716){\usebox{\plotpoint}}
\put(355,717){\usebox{\plotpoint}}
\put(353,718){\usebox{\plotpoint}}
\put(352,719){\usebox{\plotpoint}}
\put(351,720){\usebox{\plotpoint}}
\put(350,721){\usebox{\plotpoint}}
\put(349,722){\usebox{\plotpoint}}
\put(348,723){\usebox{\plotpoint}}
\put(347,724){\usebox{\plotpoint}}
\put(345,725){\usebox{\plotpoint}}
\put(344,726){\usebox{\plotpoint}}
\put(343,727){\usebox{\plotpoint}}
\put(342,728){\usebox{\plotpoint}}
\put(341,729){\usebox{\plotpoint}}
\put(340,730){\usebox{\plotpoint}}
\put(339,731){\usebox{\plotpoint}}
\put(338,732){\usebox{\plotpoint}}
\put(336,733){\usebox{\plotpoint}}
\put(335,734){\usebox{\plotpoint}}
\put(334,735){\usebox{\plotpoint}}
\put(333,736){\usebox{\plotpoint}}
\put(332,737){\usebox{\plotpoint}}
\put(331,738){\usebox{\plotpoint}}
\put(330,739){\usebox{\plotpoint}}
\put(329,740){\usebox{\plotpoint}}
\put(327,741){\usebox{\plotpoint}}
\put(326,742){\usebox{\plotpoint}}
\put(325,743){\usebox{\plotpoint}}
\put(324,744){\usebox{\plotpoint}}
\put(323,745){\usebox{\plotpoint}}
\put(321,746){\usebox{\plotpoint}}
\put(320,747){\usebox{\plotpoint}}
\put(319,748){\usebox{\plotpoint}}
\put(318,749){\usebox{\plotpoint}}
\put(317,750){\usebox{\plotpoint}}
\put(315,751){\usebox{\plotpoint}}
\put(314,752){\usebox{\plotpoint}}
\put(313,753){\usebox{\plotpoint}}
\put(312,754){\usebox{\plotpoint}}
\put(311,755){\usebox{\plotpoint}}
\put(310,756){\usebox{\plotpoint}}
\put(308,757){\usebox{\plotpoint}}
\put(307,758){\usebox{\plotpoint}}
\put(306,759){\usebox{\plotpoint}}
\put(305,760){\usebox{\plotpoint}}
\put(304,761){\usebox{\plotpoint}}
\put(303,762){\usebox{\plotpoint}}
\put(301,763){\usebox{\plotpoint}}
\put(300,764){\usebox{\plotpoint}}
\put(299,765){\usebox{\plotpoint}}
\put(298,766){\usebox{\plotpoint}}
\put(297,767){\usebox{\plotpoint}}
\put(296,768){\usebox{\plotpoint}}
\put(295,769){\usebox{\plotpoint}}
\put(293,770){\usebox{\plotpoint}}
\put(292,771){\usebox{\plotpoint}}
\put(291,772){\usebox{\plotpoint}}
\put(290,773){\usebox{\plotpoint}}
\put(289,774){\usebox{\plotpoint}}
\put(288,775){\usebox{\plotpoint}}
\put(287,776){\usebox{\plotpoint}}
\put(286,777){\usebox{\plotpoint}}
\put(284,778){\usebox{\plotpoint}}
\put(283,779){\usebox{\plotpoint}}
\put(282,780){\usebox{\plotpoint}}
\put(281,781){\usebox{\plotpoint}}
\put(280,782){\usebox{\plotpoint}}
\put(279,783){\usebox{\plotpoint}}
\put(278,784){\usebox{\plotpoint}}
\put(277,785){\usebox{\plotpoint}}
\put(1423,276){\circle{24}}
\put(1397,273){\circle{24}}
\put(1371,270){\circle{24}}
\put(1345,267){\circle{24}}
\put(1319,265){\circle{24}}
\put(1293,262){\circle{24}}
\put(1267,259){\circle{24}}
\put(1241,257){\circle{24}}
\put(1215,254){\circle{24}}
\put(1189,252){\circle{24}}
\put(1163,250){\circle{24}}
\put(1136,248){\circle{24}}
\put(1110,247){\circle{24}}
\put(1084,246){\circle{24}}
\put(1058,245){\circle{24}}
\put(1032,245){\circle{24}}
\put(1006,246){\circle{24}}
\put(980,248){\circle{24}}
\put(954,251){\circle{24}}
\put(928,256){\circle{24}}
\put(902,263){\circle{24}}
\put(876,273){\circle{24}}
\put(850,286){\circle{24}}
\put(824,302){\circle{24}}
\put(798,321){\circle{24}}
\put(772,343){\circle{24}}
\put(746,367){\circle{24}}
\put(720,392){\circle{24}}
\put(694,416){\circle{24}}
\put(668,441){\circle{24}}
\put(642,465){\circle{24}}
\put(616,489){\circle{24}}
\put(590,513){\circle{24}}
\put(564,536){\circle{24}}
\put(537,559){\circle{24}}
\put(511,582){\circle{24}}
\put(485,605){\circle{24}}
\put(459,627){\circle{24}}
\put(433,650){\circle{24}}
\put(407,673){\circle{24}}
\put(381,695){\circle{24}}
\put(355,718){\circle{24}}
\put(329,741){\circle{24}}
\put(303,763){\circle{24}}
\put(277,786){\circle{24}}
\put(276,786){\usebox{\plotpoint}}
\end{picture}

\end{figure}

\end{document}